\documentclass[11pt]{article}
\usepackage{amsmath,amsfonts,amsthm,amssymb}

\newcommand{\version}{August 30, 2006}

\font\notefont=cmsl8

\newcommand\beq{\begin{equation}}
\newcommand\eeq{\end{equation}}

\newcommand\N{{\mathbb N}}
\newcommand\C{{\mathbb C}}

\newcommand\R{{\mathbb R}}

\renewcommand\rho\varrho

\newcommand\Tr{{\rm Tr}}
\newcommand\sign{{\rm sign}}

\newtheorem{thm}{Theorem}
\newtheorem{lem}{Lemma}
\newtheorem{cor}{Corollary}

\theoremstyle{definition}
\newtheorem{rem}{Remark}

\begin{document}

\title{Lieb-Thirring inequalities for Schr\"odinger operators with
complex-valued potentials}

\author{Rupert L. Frank$^1$ \and Ari Laptev$^1$\\
Department of Mathematics\\
Royal Institute of Technology, 100 44 Stockholm, Sweden\\
\small{Email: \{\texttt{rupert, laptev}\}\texttt{@math.kth.se}}
\and
{Elliott H.~Lieb$^2$}
\and {Robert Seiringer$^3$}\\
Departments of Physics$^{2,3}$ and Mathematics$^2$\\
Princeton University, P.~O.~Box 708, Princeton, NJ 08544, USA\\
\small{Email: \{\texttt{lieb, rseiring}\}\texttt{@princeton.edu}}}
  \date{\small \version}
\maketitle

\renewcommand{\thefootnote}{$1$}
\footnotetext{Work partially supported by
the Swedish Foundation for International Cooperation in Research
and Higher Education (STINT).}
\renewcommand{\thefootnote}{$2$}
\footnotetext{Work partially
supported by U.S. National Science Foundation
grant PHY 01 39984.}
\renewcommand{\thefootnote}{$3$}
\footnotetext{Work partially
supported by U.S. National Science Foundation
grant PHY 03 53181, and by an A.P. Sloan Fellowship.\\
\copyright\, 2006 by the authors. This paper may be reproduced, in its
entirety, for non-commercial purposes.}
\renewcommand{\thefootnote}{$4$}
\renewcommand{\thefootnote}{$5$}

\begin{abstract}
Inequalities are derived for power sums of the real part and
the modulus of the eigenvalues of a Schr\"odinger operator with a
complex-valued
potential.
\end{abstract}
\bigskip

\section{Introduction}
A motivation for the present paper was a challenge posed in a lecture by
E.B.~Davies
about {\it non-}self-adjoint Schr\"odinger operators in
$L^2(\R^d)$:
\begin{equation}
H= -\Delta +V(x) \ ,
\end{equation}
where $V$ is a complex-valued potential (see also the papers \cite{AAD},
\cite{DN}).
Theorem~4 in \cite{AAD} states
that when $d=1$ every eigenvalue $\lambda$  of $H$ that does not lie on
the positive
real axis satisfies
\begin{equation}\label{davies}
|\lambda| \leq \frac14\left(\int_{\R} |V(x)| \ dx\right)^2 \ .
\end{equation}
We note that the constant $\frac 14$ in this inequality is optimal.
The  question was raised whether an estimate similar to \eqref{davies}
holds in dimension
$d\geq 2$.

While we do not answer the question directly, we have succeeded in
finding a version of the Lieb-Thirring inequality for the eigenvalue
power sums (Riesz means) that holds for this non-self-adjoint
operator. Since little is  known about non-self-adjoint operators
relative to self-adjoint operators, our results may be worth
recording. The proofs are easy, but not entirely obvious.

We denote by $\lambda_j$, \ $j=1,2,3,\ldots\,$, a listing of the (countably
many) eigenvalues of $H$ in the cut plane $\C\setminus[0,\infty)$,
repeated according to their \emph{algebraic} multiplicities.  An
eigenvalue is a solution to the equation $H\psi = \lambda \psi$ for
some $\psi \in L^2$.  A given number $\lambda \in \C$ may occur
several times in this list of eigenvalues according to the dimension
of the generalized eigenspace $\{\psi: (H-\lambda)^k\psi=0 \text{ for
  some } k\in\N\}$, which is called the algebraic multiplicity.  In
principle a generalized eigenspace could have infinite dimension,
but, as we shall see, this will not occur in the situations considered
here.

Note that the dimension of a generalized eigenspace may be strictly
larger than the number of linearly independent solutions of $H\psi =
\lambda \psi$, i.e., the geometric multiplicity of $\lambda$.  The
algebraic multiplicity is known to be finite for sufficiently decaying
potentials as a consequence of Weyl's theorem, but we do not need this
fact in our proof; a simple corollary of our theorems is that the
multiplicity is automatically finite when the appropriate power of the
potential is integrable.

It is a pleasure to acknowledge some very fruitful discussions with
Prof. E.B. Davies about this paper, especially with regard to the
question of multiplicities. Our original version was formulated in
terms of geometric multiplicities instead of algebraic multiplicities
because we needed to use the actual eigenfunctions of $H$, and these
exist only with geometric multiplicity. He pointed out that it is only
necessary in our proof to have basis functions in the generalized
eigenspace with eigenvalue $\lambda$ such that $(\phi, H\phi) =
\lambda (\phi,\phi) = \lambda \|\phi\|^2$.

Before stating our main results let us recall the standard
Lieb-Thirring inequalities (see \cite{LT} and also the survey
\cite{LaW}). For real-valued potentials $V$ one has the bound
\begin{equation}\label{lt}
\sum_j (\lambda_j)_-^\gamma
\leq L_{\gamma,d} \int_{\R^d} V(x)_-^{\gamma+d/2}\,dx \,
\end{equation}
provided $\gamma\geq\frac12$ if $d=1$, $\gamma>0$ if $d=2$ and
$\gamma\geq 0$ if $d\geq3$. (Here and in the sequel $t_-:=\max\{ 0,
-t\}$ denotes the negative part of $t$.)  By $L_{\gamma,d}$ we will
always mean the \emph{sharp} constant in \eqref{lt} (which at present
is only known for $\gamma=\frac12$ if $d=1$ and for
$\gamma\geq\frac32$ if $d\geq1$, see \cite{LaW}).

For general, complex-valued potentials we shall prove

\begin{thm}[Eigenvalue sums]   \label{evsums}
Let $d\geq 1$ and $\gamma\geq 1$.
\begin{enumerate}
\item
For eigenvalues with non-positive real parts
\begin{equation}\label{evsums1}
\sum_{\Re\lambda_j<0} (-\Re\lambda_j)^\gamma
\leq L_{\gamma,d} \int_{\R^d} (\Re V(x))_-^{\gamma+d/2}\,dx.
\end{equation}
\item
If $\varkappa>0$, then for eigenvalues outside the cone $\{|\Im
z|<\varkappa\,\Re z\}$,
\begin{equation}\label{evsums2}
\sum_{|\Im\lambda_j|\geq\varkappa\,\Re\lambda_j} |\lambda_j|^\gamma
\leq C_{\gamma,d}(\varkappa) \int_{\R^d} |V(x)|^{\gamma+d/2}\,dx.
\end{equation}
\end{enumerate}
Here $L_{\gamma,d}$ is the same as the constant in \eqref{lt} and
\linebreak
$C_{\gamma,d}(\varkappa) = 2^{1+\gamma/2+d/4}
\left(1+\frac{2}{\varkappa}\right)^{\gamma+d/2} L_{\gamma,d}$.
\end{thm}

As a consequence we obtain

\begin{cor} \label{evsumscor}
Let $d\geq 1$ and $\gamma\geq 1$.
\begin{enumerate}
\item
For eigenvalues with non-positive real parts
\begin{equation}\label{evsums1cor}
\sum_{\Re\lambda_j<0} |\lambda_j|^\gamma
\leq C_{\gamma,d} \int_{\R^d} |V(x)|^{\gamma+d/2}\,dx.
\end{equation}
\item
If $\varkappa>0$, then for eigenvalues inside the cone $\{|\Im
z|\leq-\varkappa\,\Re z\}$
\begin{equation}\label{evsums2cor}
\sum_{|\Im\lambda_j|\leq-\varkappa\,\Re\lambda_j} |\lambda_j|^\gamma
\leq L_{\gamma,d}(\varkappa) \int_{\R^d} (\Re V(x))_-^{\gamma+d/2}\,dx.
\end{equation}
Here $C_{\gamma,d} = 2^{1+\gamma/2+d/4} L_{\gamma,d}$ and
$L_{\gamma,d}(\varkappa) = (1+\varkappa) L_{\gamma,d}$.
\end{enumerate}
\end{cor}

It is natural to conjecture that the estimates in Theorem \ref{evsums} and
Corollary~\ref{evsumscor} hold for all values of $\gamma$ for which
\eqref{lt} holds, and not only for $\gamma\geq 1$.

The proof below shows that $|V(x)|$ in the bounds \eqref{evsums2} and
\eqref{evsums1cor} can actually be replaced by $\frac 1{\sqrt2} \left((\Re
V(x))_- +|\Im V(x)|\right)$.

\begin{rem} We can replace $-\Delta $ in $H$ by $(i\nabla +A(x))^2$, where
  $A$ is an arbitrary, real vector-field. This replacement is valid
  for the usual (self-adjoint) Lieb-Thirring inequality \eqref{lt},
  and so it is valid here because we use only the self-adjoint
  Lieb-Thirring inequality in our proof of the theorem. If $d=1$ or if
  $\gamma\geq\frac32$ the constant in \eqref{lt} (and hence in Theorem
  \ref{evsums} and Corollary \ref{evsumscor}) remains the same as in
  the case $A=0$. In general it is not known whether the constant
$L_{\gamma,d}$ in
\eqref{lt}
   has to be increased when the $A$ is added. It is a fact, however,
  that all known proofs of the Lieb-Thirring inequality (without the,
  as yet unknown, sharp constant) do not require an increase in
  the constant.
\end{rem}

\begin{rem} We can also replace $-\Delta $ in $H$ by any operator for which
  Lieb-Thirring bounds for real-valued potentials hold (but making the
  appropriate change in the exponent of $V$ on the right side of the
  inequalities). For example, we can replace $-\Delta $ in $H$ by the
  ``relativistic'' operator $|i\nabla +A(x)| $, in
  which case $\gamma +d/2$ has to be replaced by $\gamma +d$.
\end{rem}

We now state
bounds on single eigenvalues. Let us denote by
$L^1_{\gamma,d}$ the sharp constant in the inequality
\begin{equation}\label{ltone}
\left(\inf {\rm spec}  (-\Delta+V)\right)_-^\gamma
\leq L^1_{\gamma,d} \int_{\R^d} V(x)_-^{\gamma+d/2}\,dx
\end{equation}
for real-valued potentials $V$. This estimate holds under the same
condition on $\gamma$ as \eqref{lt} and one has, of course,
$L^1_{\gamma,d}\leq L_{\gamma,d}$. The sharp value of $L^1_{\gamma,d}$
is known for $\gamma\geq\frac12$ if $d=1$ and for $\gamma=0$ if
$d\geq3$. Note that $L^1_{\gamma,d}$ plays a role in the Lieb-Thirring
conjecture, see \cite{LT}.

\begin{thm}[Bounds on single eigenvalues]   \label{oneev}
  Let $\gamma\geq\frac12$ if $d=1$, $\gamma>0$ if $d= 2$ and
  $\gamma\geq0$ if $d\geq 3$.
\begin{enumerate}
\item
For any eigenvalue with non-positive real part
\begin{equation}\label{smallestev1}
(-\Re\lambda_j)^\gamma \leq L^1_{\gamma,d} \int_{\R^d} (\Re
V(x))_-^{\gamma+d/2}\,dx
\end{equation}
and
\begin{equation}\label{smallestev2}
|\lambda_j|^\gamma \leq C^1_{\gamma,d} \int_{\R^d} |V(x)|^{\gamma+d/2}\,dx.
\end{equation}
\item
For any eigenvalue with non-negative real part
\begin{equation}\label{smallestev3}
|\lambda_j|^\gamma
\leq
C^1_{\gamma,d}\left(1+\frac{2\Re\lambda_j}{|\Im\lambda_j|}\right)^{\gamma+d/2}
\int_{\R^d} |V(x)|^{\gamma+d/2}\,dx.
\end{equation}
\end{enumerate}
Here $L_{\gamma,d}^1$ is the same as the constant in \eqref{ltone} and
$C_{\gamma,d}^1 = 2^{\gamma/2+d/4} L_{\gamma,d}^1$.
\end{thm}

\begin{rem}
  This theorem yields a region in $\C$ in which there are no
  eigenvalues. This region is far from optimal; in particular, it does
  not approach the positive real axis as $\lambda$ gets large. The
  paper \cite{DN} of Davies--Nath has a much better result for $d=1$.
\end{rem}


\section{Proof of Theorem \ref{evsums}}

The core of Theorem \ref{evsums} is contained in

\begin{lem} Let $\lambda_1, \ \lambda_2, \dots , \lambda_N$ be an
  arbitrary finite family of eigenvalues of $H$ in $\C\setminus
  [0,\infty)$.  (In the case of algebraic multiplicity $k>1$
  a given number $\lambda\in \C$ might occur several times in our
  family, but no more than $k$ times.)  Then, for any $\alpha\in\R$ and
$\gamma\geq 1$,
\begin{equation}\label{lemma1}
\sum_{j =1} ^N (\Re\lambda_j+\alpha\Im\lambda_j)_-^\gamma
\leq \Tr(-\Delta+\Re V+\alpha \Im V)_-^\gamma.
\end{equation}
\end{lem}

\begin{proof} We begin with the case $\gamma=1$. Let $\alpha\in\R$. By
removing some of the $\lambda_j$ we can assume without loss of generality
that $-\Re\lambda_j -\alpha\Im\lambda_j>0$ for all $1\leq j\leq N$.

Special attention must be given to a number $\lambda$ that occurs
several times in our list owing to an algebraic multiplicity $>1$.
Suppose that this $\lambda $ occurs $k$ times (while the algebraic
multiplicity is $\geq k$). We can always find orthonormal functions
$\varphi_1, ..., \varphi_k$ in the invariant subspace belonging to $\lambda$
such that $H\varphi_1=\lambda \varphi_1$ and
\begin{equation}\label{basis}
H\varphi_j = \lambda_j\varphi_j + \sum_{k< j} \alpha_{kj}\varphi_k.
\end{equation}
This is the upper triangular representation familiar from elementary
linear algebra.

The collection of all the $\varphi_j$ for the different eigenvalues in
our family yields $N$ linearly independent functions, which we denote
by $\psi_j$, $j=1,\dots,N$.  We introduce the function of $N$
variables in $\R^{d}$
\begin{equation*}  \label{det}
\Psi (x_1,\dots,x_N) := \det\left(\psi_j(x_k)\right),
\qquad (x_1,\dots,x_N)\in\R^{dN}.
\end{equation*}
The linear independence of the $\psi_j$ implies that $\Psi\not\equiv 0$.
An easy
calculation using \eqref{basis} shows that
\begin{equation*}\label{lemmaproof}
\sum_{j=1}^N \lambda_j \int_{\R^{dN}} |\Psi|^2 \,dx_1\dots dx_N
= \sum_{j=1}^N \int_{\R^{dN}} \left( |\nabla_j \Psi|^2 +
V(x_j)|\Psi|^2\right)
\,dx_1\dots dx_N,
\end{equation*}
where $\nabla_j$ denotes the gradient with respect to the variable $x_j$.
Taking the
real part in this relation we find that
\begin{equation*}
\|\Psi\|^2 \sum_{j=1}^N\Re\lambda_j    = (\Psi, H^{(N)}\Psi),
\end{equation*}
where $H^{(N)}:=\sum_{j=1}^N (-\Delta_j+\Re V(x_j))$ acting on
antisymmetric functions in $L^2(\R^{dN})$.  A similar equality holds
for $\Im \lambda_j$. Adding these two equations we have that
\begin{equation} \label{alphasum}
\|\Psi\|^2  \sum_{j=1} ^N(\Re\lambda_j+\alpha \Im\lambda_j)   = (\Psi,\tilde
H^{(N)}\Psi) \ ,
\end{equation}
where now $\tilde H^{(N)}:=\sum_{j=1}^N (-\Delta_j+\Re V(x_j)   +\alpha
\Im V(x_j) )$.

The variational principle together with \eqref{alphasum} implies
\begin{equation*}
\sum_{j=1} ^N
(\Re\lambda_j + \alpha \Im\lambda_j) \geq \inf {\rm spec} \left(\tilde
H^{(N)}\right) \geq -\Tr(-\Delta+\Re V +\alpha \Im V)_-.
\end{equation*}
This proves \eqref{lemma1} in the case $\gamma=1$.

Now we reduce the case $\gamma>1$ to the previous one following an idea
of Aizenman-Lieb in  \cite{LA}. There is a constant $C_\gamma$ such that
\begin{equation}\label{beta}
C_\gamma s_-^\gamma = \int_0^\infty t^{\gamma-2} (s+t)_- \, dt.
\end{equation}
(Indeed, $C_\gamma$ can be expressed in terms of the beta function, but we
will not need this.) Hence
\begin{equation*}
C_\gamma \sum_{j=1}^N (\Re\lambda_j+\alpha\Im\lambda_j)_-^\gamma
= \int_0^\infty t^{\gamma-2} \sum_{j=1}^N
(\Re\lambda_j(t)+\alpha\Im\lambda_j(t))_-\,dt
\end{equation*}
where $\lambda_j(t):=\lambda_j+t$. The numbers $\lambda_j(t)$ are the
eigenvalues of the operator $-\Delta+V_t$, $V_t(x):=V(x)+t$. Applying the
result for $\gamma=1$ that we have already proved we get
\begin{equation*}
\sum_{j=1}^N (\Re\lambda_j(t)+\alpha\Im\lambda_j(t))_-
\leq \Tr(-\Delta+\Re V_t+\alpha \Im V_t)_-
= \Tr(-\Delta+\Re V+\alpha \Im V +t)_-.
\end{equation*}
Using \eqref{beta} once more we conclude that
\begin{equation*}
\begin{split}
C_\gamma \sum_{j=1}^N (\Re\lambda_j+\alpha\Im\lambda_j)_-^\gamma
& \leq \int_0^\infty t^{\gamma-2} \Tr(-\Delta+\Re V+\alpha \Im V +t)_-
\,dt \\
& = C_\gamma \, \Tr(-\Delta+\Re V+\alpha \Im V)_-^\gamma,
\end{split}
\end{equation*}
as claimed.
\end{proof}

Now everything is in place for the

\begin{proof}[Proof of Theorem \ref{evsums}] The
estimate \eqref{evsums1}
follows immediately from \eqref{lemma1} with $\alpha=0$ and \eqref{lt}.

To obtain the estimate \eqref{evsums2} we apply \eqref{lemma1} with
$\alpha=1+\frac 2\varkappa$, considering only those eigenvalues with
$\Im\lambda_j \leq 0$ and $\varkappa\,\Re\lambda_j\leq -\Im\lambda_j$.
(If there are infinitely many eigenvalues we consider a finite subset
and pass to the limit.) We get
\begin{equation*}
\sum_{\substack{ \Im\lambda_j \leq 0,\\ \varkappa\,\Re\lambda_j\leq
-\Im\lambda_j}}
\left(-\Re\lambda_j-\left(1+\frac 2\varkappa\right)\Im\lambda_j\right)^\gamma
\leq \Tr\left(-\Delta+\Re V+\left(1+\frac 2\varkappa\right)\Im
V\right)_-^\gamma.
\end{equation*}
Now replace $V$ in this inequality by its complex conjugate $\overline{V}$
and note
that the
eigenvalues
of the operator $-\Delta+\overline{V}$ are $\overline{\lambda_j}$. Hence
\begin{equation*}
\sum_{\substack{\Im\lambda_j \geq 0,\\ \varkappa\,\Re\lambda_j\leq
\Im\lambda_j}}
\left(-\Re\lambda_j+\left(1+\frac 2\varkappa\right)\Im\lambda_j\right)^\gamma
\leq \Tr\left(-\Delta+\Re V-\left(1+\frac 2\varkappa\right)\Im
V\right)_-^\gamma.
\end{equation*}
Adding the two previous relations yields
\begin{equation*}
\begin{split}
& \sum_{\varkappa\,\Re\lambda_j\leq |\Im\lambda_j|}
\left(-\Re\lambda_j+\left(1+\frac
2\varkappa\right)|\Im\lambda_j|\right)^\gamma \\
& \qquad\leq \Tr\left(-\Delta+\Re V+\left(1+\frac 2\varkappa\right)\Im
V\right)_-^\gamma \\
& \qquad\qquad +
\Tr\left(-\Delta+\Re V-\left(1+\frac 2\varkappa\right)\Im V\right)_-^\gamma.
\end{split}
\end{equation*}
Now \eqref{evsums2} follows from \eqref{lt} by means of the elementary
inequality $\sqrt{a^2+b^2}\leq a+b \leq \sqrt 2\sqrt{a^2+b^2}$ for
$a,b\geq 0$ and the bound
\begin{equation*}
-\Re\lambda_j+\left(1+\frac 2\varkappa\right)|\Im\lambda_j|\geq
|\Re\lambda_j|+|\Im\lambda_j|
\end{equation*}
provided $\varkappa\,\Re\lambda_j\leq |\Im\lambda_j|$.
\end{proof}

\begin{proof}[Proof of Corollary \ref{evsumscor}]
  The estimate \eqref{evsums1cor} follows from \eqref{evsums2} by
  letting $\varkappa\to\infty$, and the estimate \eqref{evsums2cor}
  follows by noting that $|\Re\lambda_j|+|\Im\lambda_j|\leq
  (1+\varkappa)|\Re\lambda_j|$ provided $-\varkappa\,\Re\lambda_j\geq
  |\Im\lambda_j|$.
\end{proof}


\section{Proof of Theorem \ref{oneev}}

We proceed similarly as in the proof of the previous theorem. Let
$\psi_j$ be an eigenfunction corresponding to an eigenvalue
$\lambda_j$.  Considering the real and imaginary parts of the equation
\begin{equation*}
\int\left(|\nabla \psi_j|^2 +V|\psi_j|^2\right)\, dx = \lambda_j \int
|\psi_j|^2\,dx
\end{equation*}
we find that for any $\alpha\in\R$
\begin{equation*}
\int\left(|\nabla \psi_j|^2 +\Re V|\psi_j|^2+\alpha\Im V|\psi_j|^2
\right)\, dx
= (\Re\lambda_j+\alpha\Im\lambda_j) \int |\psi_j|^2\,dx.
\end{equation*}
The variational principle implies
\begin{equation*}
\inf {\rm spec} (-\Delta+\Re V+\alpha\Im V) \leq
\Re\lambda_j+\alpha\Im\lambda_j.
\end{equation*}
The estimates \eqref{smallestev1}, \eqref{smallestev2} for eigenvalues
with non-positive real part follow now with the choices $\alpha=0$ and
$\alpha=-\sign\Im\lambda_j$, respectively, from \eqref{ltone}. Similarly,
\eqref{smallestev3} for eigenvalues with non-negative real part is
obtained by the choice
$\alpha
=(-\sign\Im\lambda_j)\left(1+\frac{2\Re\lambda_j}{|\Im\lambda_j|}\right)$.



\begin{thebibliography}{9}

\bibitem{AAD}
A.A. Abramov, A. Aslanyan and E.B. Davies, {\it Bounds on complex
eigenvalues and
resonances},
Jour. Phys. A {\bf 34}, 57-72 (2001).

\bibitem{DN}
E.B. Davies and J. Nath, {\it Schr\"odinger operators with slowly decaying
potentials},
J. Comput. Appl. Math. {\bf 148}, no. 1, 1--28 (2002).

\bibitem{LA}  M. Aizenman and E.H. Lieb, {\it On Semi-Classical
Bounds for Eigenvalues of Schr\"odinger Operators}, Phys. Lett. {\bf 66A},
427-429
(1978).

\bibitem{LaW} A. Laptev and T. Weidl, {\it Recent results
  on Lieb-Thirring inequalities}, Journ\'ees "\'Equations aux D\'eriv\'ees
  Partielles" (La Chapelle sur Erdre, 2000), Exp. No. XX,
  Univ. Nantes, Nantes (2000).

\bibitem{LT}
E.H. Lieb and W. Thirring, {\it  Inequalities for the Moments of the
Eigenvalues of the Schr\"odinger Hamiltonian and Their Relation to Sobolev
Inequalities}, in {\it Studies in Mathematical Physics}, E. Lieb, B. Simon,
A. Wightman eds., Princeton University Press, 269-303 (1976).




\end{thebibliography}
\end{document}